\documentclass{PoS}

\usepackage{amsfonts}
\usepackage{amsmath}
\usepackage{amssymb}
\usepackage{latexsym}
\usepackage{epsfig}
\usepackage{array}
\usepackage{pifont}
\usepackage{axodraw}
\usepackage{citesort}
\usepackage{braket}
\newcommand{\msbar}{\overline{\mbox{{\sc ms}}}}
\newcommand{\VA}{\braket{A^2}}

\newcommand{\beq}{\begin{eqnarray}}
\newcommand{\eeq}{\end{eqnarray}}

\newcommand{\be}{\begin{equation}}
\newcommand{\ee}{\end{equation}}
\newcommand{\lwrsim}{\raise0.3ex\hbox{$<$\kern-0.75em\raise-1.1ex\hbox{$\sim$}}}

\def\C2#1#2{({\cal C}_2)_{#1}^{#2}}

\def\eq#1{Eq.~(\ref{#1})}
\def\altura#1{\rule[0cm]{0cm}{#1cm}}

\newcommand{\ghThreeOneRS}{\begin{picture}(150,45)(0,0)
\SetWidth{1.2}
\DashArrowLine(12.5,0)(37.5,0){5}
\DashArrowLine(37.5,0)(112.5,0){5}
\DashArrowLine(112.5,0)(137.5,0){5}
\SetWidth{1}
\Vertex(37.5,0){2}
\Vertex(112.5,0){2}
\Vertex(40,31){2}
\Gluon(37.5,0)(37.5,48){-4}{4}
\GlueArc(67.5,0)(45,70,135){-4}{4}
\GlueArc(67.5,0)(45,0,70){-4}{5}
\CCirc(78,40){10}{Black}{Blue}
\Text(135,5)[]{q}
\Text(20,5)[]{k}
\Text(25,47)[]{q-k}
\end{picture}}
\newcommand{\ghThreeOneRSs}{\begin{picture}(150,45)(0,0)
\SetWidth{1.2}
\DashArrowLine(12.5,0)(37.5,0){5}
\DashArrowLine(37.5,0)(112.5,0){5}
\DashArrowLine(112.5,0)(137.5,0){5}
\SetWidth{1}
\Vertex(37.5,0){2}
\Vertex(112.5,0){2}
\Vertex(109,34){2}
\Gluon(112.5,0)(112.5,48){4}{4}
\GlueArc(82.5,0)(45,53,100){-4}{3}
\GlueArc(82.5,0)(45,100,180){-4}{5}
\CCirc(70,40){10}{Black}{Blue}
\Text(135,5)[]{q}
\Text(20,5)[]{k}
\Text(125,47)[]{q-k}
\end{picture}}
\newcommand{\ghThreeTwo}{\begin{picture}(120,40)(0,0)
\SetWidth{1.2}
\DashArrowLine(10,20)(30,20){5}
\DashArrowLine(30,20)(90,20){5}
\DashArrowLine(90,20)(110,20){5}
\SetWidth{1}
\Vertex(30,20){2}
\Vertex(90,20){2}
\Gluon(60,20)(60,-10){4}{3}
\GlueArc(60,20)(30,0,75){-4}{4}
\GlueArc(60,20)(30,105,180){-4}{4}
\CCirc(60,50){10}{Black}{Blue}
\end{picture}}
\newcommand{\ghThreeThree}{\begin{picture}(130,25)(0,0)
\SetWidth{1.2}
\DashArrowLine(10,0)(30,0){5}
\DashArrowLine(30,0)(90,0){5}
\DashArrowLine(90,0)(120,0){5}
\SetWidth{1}
\Vertex(30,0){2}
\Vertex(90,0){2}
\Vertex(112,0){2}
\Gluon(112,0)(112,40){4}{4}
\GlueArc(60,0)(30,0,75){-4}{4}
\GlueArc(60,0)(30,105,180){-4}{4}
\CCirc(60,30){10}{Black}{Blue}
\end{picture}}

\newcommand{\ghThreeThreeRS}{\begin{picture}(130,25)(0,0)
\SetWidth{1.2}
\DashArrowLine(10,0)(30,0){5}
\DashArrowLine(30,0)(90,0){5}
\DashArrowLine(90,0)(120,0){5}
\SetWidth{1}
\Vertex(30,0){2}
\Vertex(90,0){2}
\Vertex(112,0){2}
\Gluon(30,0)(30,40){4}{4}
\GlueArc(82,0)(30,0,75){-4}{4}
\GlueArc(82,0)(30,105,180){-4}{4}
\CCirc(82,30){10}{Black}{Blue}
\Text(10,7)[]{k}
\Text(15,35)[]{q-k}
\Text(125,7)[]{q}
\end{picture}}
\newcommand{\ghThreeFour}{\begin{picture}(120,40)(0,0)
\SetWidth{1.2}
\DashArrowLine(110,-10)(110,20){5}
\DashArrowLine(110,20)(110,50){5}
\SetWidth{1}
\Gluon(10,20)(30,20){4}{2}
\Gluon(30,20)(90,20){4}{6}
\Gluon(90,20)(110,20){4}{2}
\Vertex(30,20){2}
\Vertex(90,20){2}
\Vertex(110,20){2}
\GlueArc(60,20)(30,0,75){-4}{4}
\GlueArc(60,20)(30,105,180){-4}{4}
\CCirc(60,50){10}{Black}{Blue}
\end{picture}}

\newcommand{\ghThreeFive}{\begin{picture}(120,40)(0,0)
\SetWidth{1.2}
\DashArrowLine(110,-10)(110,10){5}
\DashArrowLine(110,10)(110,50){5}
\SetWidth{1}
\Gluon(10,10)(60,10){4}{6}
\Gluon(60,10)(110,10){4}{6}
\Vertex(60,10){2}
\Vertex(110,10){2}
\GlueArc(60,35)(25,105,270){4}{7}
\GlueArc(60,35)(25,-90,75){4}{7}
\CCirc(60,60){10}{Black}{Blue}
\end{picture}}

\title{The dimension-two gluon condensate, the ghost-gluon vertex 
and the Taylor theorem}

\ShortTitle{The dimension-two gluon condensate}

\author{\speaker{J.~Rodr\'iguez-Quintero}\thanks{This talk is mainly based on the work 
developped in collaboration with: Ph.~Boucaud, D. Dudal, J.P.~Leroy \& O.~P\`ene.}
\\
        Dpto. F\'isica Aplicada, Fac. CCEE, Univ. Huelva; 21071 Huelva; Spain \\
        E-mail: \email{jose.rodriguez@dfaie.uhu.es}}

\abstract{
We study the genuine non-perturbative corrections
to the Landau gauge ghost-gluon vertex in terms of the non-vanishing dimension-two
gluon condensate, and prove these corrections to give account of current SU(2) lattice data
for the vertex with different kinematical configurations in the domain of intermediate
momenta, roughly above 2-3 GeV. Based on this OPE analysis, we also present
a simple model for the vertex, in acceptable agreement with the lattice data also in the
IR domain. The necessity of a non-vanishing dimension-two gluon condensate will be also 
investigated through the analysis of the running coupling defined by the ghost-gluon vertex 
in Taylor kinematics.}

\FullConference{International Workshop on QCD Green's Functions, Confinement and Phenomenology,\\
		September 05-09, 2011\\
		Trento Italy}

\begin{document}

\section{Introduction}

The infrared properties of the Landau-gauge QCD Green functions have been triggering many studies
in the last few years, mainly involving both
lattice (see for instance Refs.~\cite{Cucchieri:2007rg,Costa:2010pp,Bogolubsky:2007bw}
) 
and continuum approaches (see for instance Refs.~\cite{Aguilar:2008xm,Binosi:2009qm,Boucaud:2008ky,Boucaud:2008ji,Boucaud:2010gr,RodriguezQuintero:2010wy,Fischer:2008uz,Fischer:2009tn,Alkofer:2008jy} using Dyson-Schwinger equations (DSE), \cite{Dudal:2005na,Dudal:2007cw,Dudal:2008sp,Dudal:2010tf} using the refined Gribov-Zwanziger formalism, \cite{Tissier:2010ts} 
using the Curci-Ferrari model as an effective description) or based 
on the infrared mapping of $\lambda \phi^4$ and Yang-Mills teories~\cite{Frasca:2007uz}.

Most of the DSE analysis take advantage of approximating the ghost-gluon vertex by a constant when truncating
the infinite tower of the relevant equations. To be more precise, only the behaviour of the involved transverse
form factor needs to be approximated by a constant for the purpose of truncating the ghost
propagator DSE (see also the recent paper \cite{Pennington:2011xs}).
Most of the ammo for the approximation of the ghost-gluon vertex and GPDSE truncation is mainly provided by
the Taylor theorem which is widely known as a non-renormalization one~\cite{Taylor:1971ff}.
This claims that, in the particular kinematical configurations defined by a vanishing incoming ghost-momentum,
no non-zero radiative correction survives for the Landau-gauge ghost-gluon vertex, which takes thus its tree-level
expression at any perturbative order~\cite{Taylor:1971ff}. This statement entails two important consequences: that (i)
the bare ghost-gluon vertex, defined by
\beq\label{vertMinko}
\Gamma^{abc}_{\mu}(-q,k;q-k) \ = \ - g_0 f^{abc}  \
\left( \altura{0.5} q_\mu H_1(q,k) + (q-k)_\mu H_2(q,k) \right) \ ,
\eeq
is UV-finite for any kinematical configuration;
and that (ii) in the specific MOM scheme where the renormalization point is taken with a vanishing
incoming ghost momentum (which we named the Taylor scheme --T-scheme-- in \cite{Boucaud:2008gn})
the renormalization constant for
the ghost-gluon vertex\footnote{It  can also be  straightforwardly concluded that the Landau-gauge ghost-gluon vertex
renormalization constant, $\widetilde{Z}_1$, is exactly 1 in the $\msbar$ scheme.}  is exactly~1.
This allows for a precise lattice determination of the strong running coupling defined in 
T-scheme~\footnote{It has been also used to
define an effective charge that could be applied for phenomenological purposes~\cite{Aguilar:2009nf}.} 
that can be confronted to perturbative prediction in order to estimate 
$\Lambda_{\overline{\rm MS}}$~\cite{Boucaud:2008gn,Blossier:2010ky,Sternbeck:2010xu}. 
The transverse character of the gluon propagator, in the Landau gauge GPDSE kernel, allows to project out the $H_2$-form factor from
expression \eqref{vertMinko}. Therefore, we have chosen to call the surviving piece, i.e.~$H_1$, the transverse form factor. Extending this
nomenclature, we call $H_2$ the longitudinal vertex form factor.

A precise determination of the ghost dressing function obtained by solving the GPDSE,
to be confronted for instance with lattice data, thus requires a correspondingly precise knowledge of the transverse form factor $H_1$.
To investigate the non-perturbative structure of that form factor 
was one of main goals of ref.~\cite{Boucaud:2011eh}, which we will pay attention to in this contribution. 
That was accomplished, as will be also shown here, by studying the non-perturbative OPE corrections to 
the perturbative form factor. The non-vanishing dimension-two gluon condensate, $\VA$, plays a crucial 
r\"ole for the leading non-perturbative contribution. This condensate was proved not to be 
neglegible when studying the running of the QCD Green 
functions~\cite{Boucaud:2000nd,Boucaud:2001qz,Boucaud:2001st,DeSoto:2001qx,Boucaud:2005xn,Boucaud:2008gn,Blossier:2010ky,Blossier:2010vt}, 
but very specially from the analysis of the running of the T-scheme coupling computed by means of lattice simulations with 
$2+1+1$ twisted-mass dynamical flavours in ref.~\cite{Blossier:2011tf}, as will be also shown here.

\section{The OPE for the T-scheme running coupling and $\VA$}
\label{sec:VA}

\begin{figure}[h]
  \begin{center}
    \begin{tabular}{cc}
    \includegraphics[width=7.3cm]{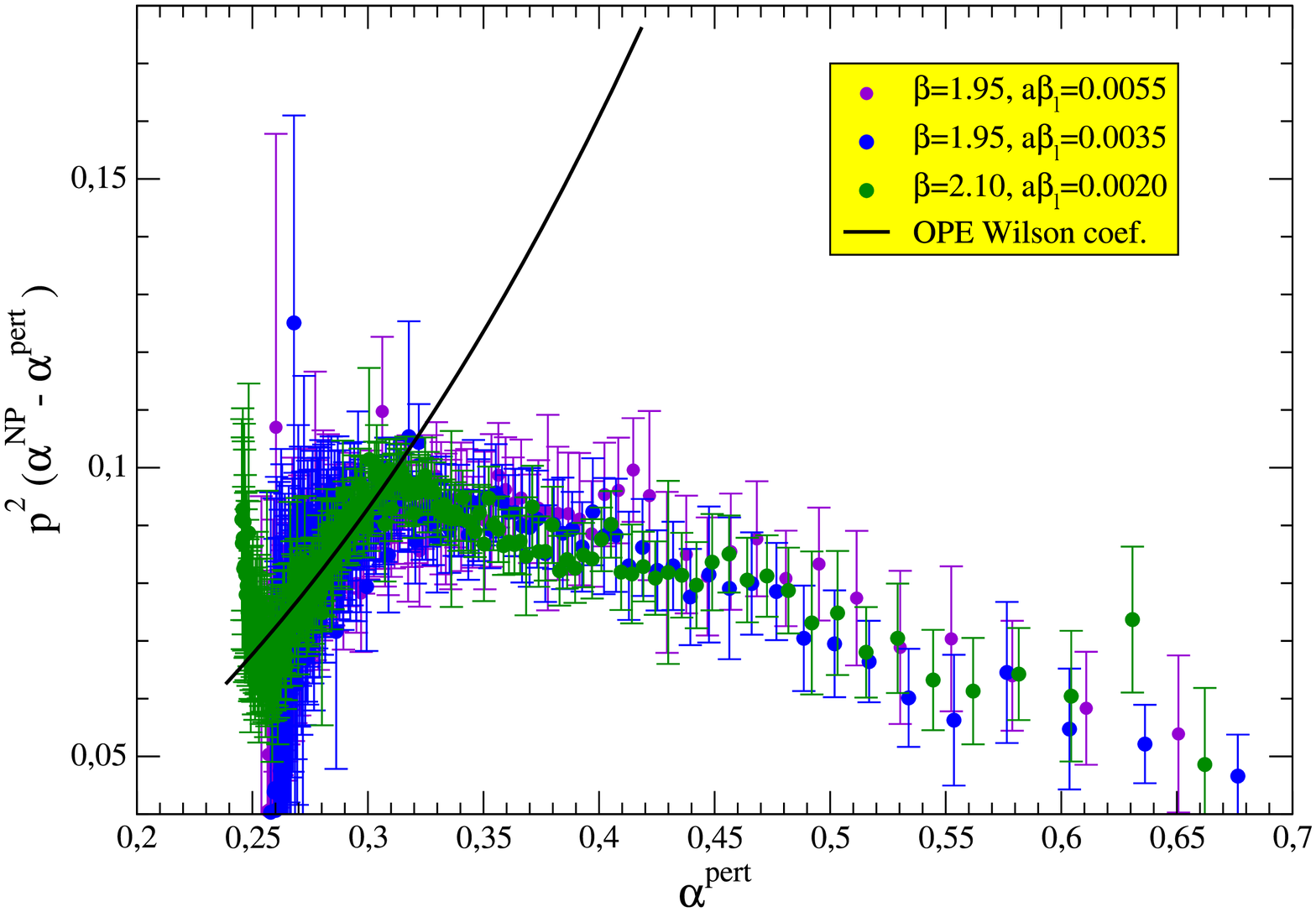} &
    \includegraphics[width=7cm]{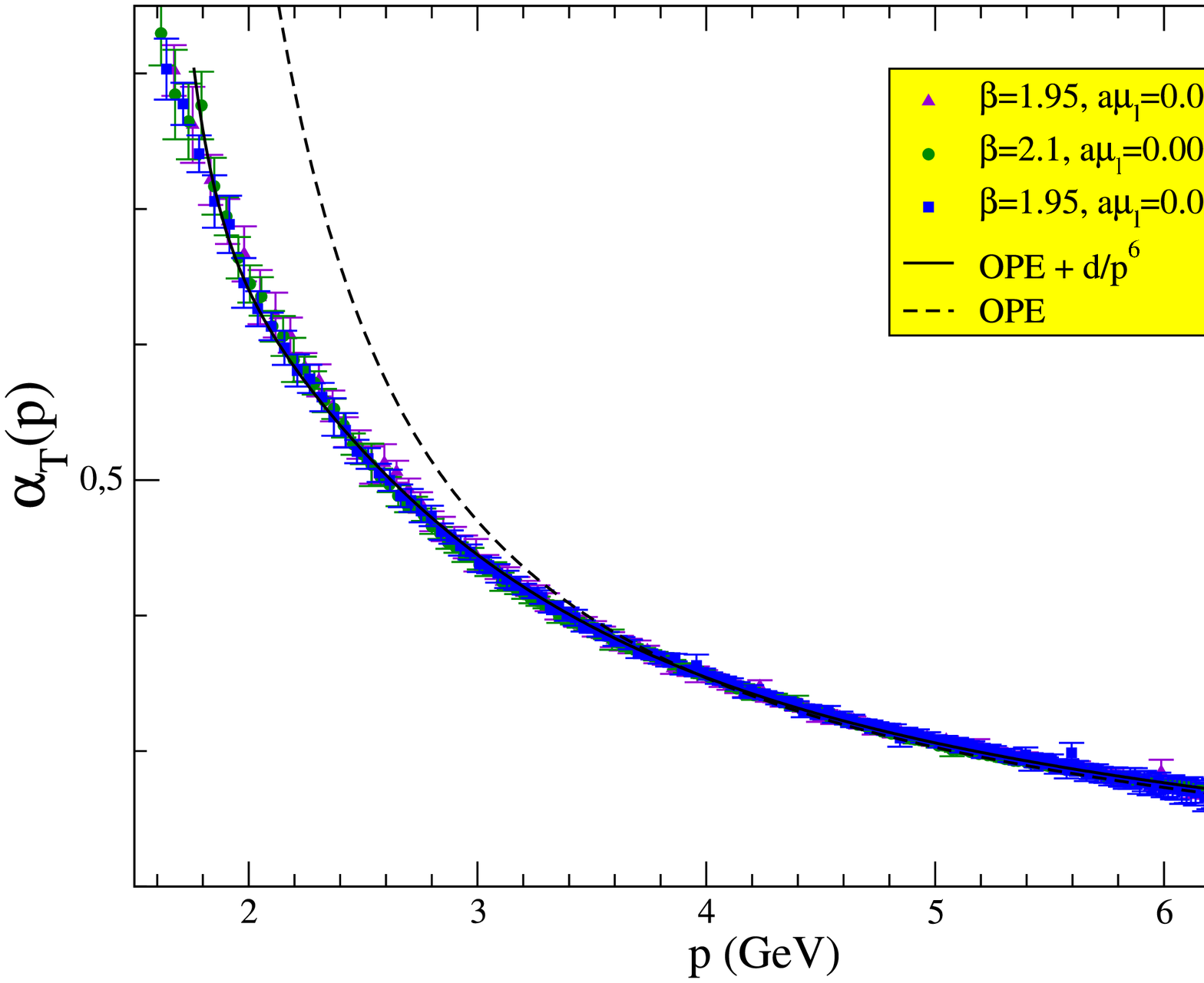}
    \end{tabular}
  \end{center}
\caption{
\small (Left) deviation from the lattice data with respect to the prediction of the 
four-loop perturbative theory, with the best-fit of $\Lambda_T$, plotted in terms of the 
perturbative running; the solid line shows the leading non-perturbative OPE prediction, 
Eq.~(2.2)'s r.h.s. (Right) The strong running coupling in Taylor scheme defined by 
Eq.~(2.1)  
obtained over a large momentum range from lattice QCD in ref.~\cite{Blossier:2011tf}. 
The dotted line stands here for the best-fit with Eq.~(2.2), 
while the solid one includes a higher-order power correction effectively 
behaving as $\sim 1/p^6$.}
\label{fig:plotalphaTPhi}
\end{figure}

A very recent analysis~\cite{Blossier:2011tf} of the strong coupling in T-scheme, 
\beq\label{alpha}
\alpha_T(q^2) \ = \ \lim_{\Lambda \to 0} \frac{g_0^2(\Lambda^2)}{4\pi} \ G(q^2,\Lambda^2) F^2(q^2,\Lambda^2) 
\overbrace{\left( H_1(q,0) + H_2(q,0) \right)}^{\displaystyle  =1 \ \{\mbox{\rm Taylor theor.} \} } \ ,
\eeq
where the bare ghost and gluon dressing functions, $G$ and $F$, were computed 
from lattice simulations with $2+1+1$ twisted-mass dynamical flavours, provided with a 
very strong evidence for the necessity of applying non-perturbative corrections to 
account for the running of the coupling,
\beq\label{alphahNP}
\alpha_T(\mu^2)
\ = \
\alpha^{\rm pert}_T(\mu^2)
\ 
\left( 
 1 + \frac{9}{\mu^2} \
R\left(\alpha^{\rm pert}_T(\mu^2),\alpha^{\rm pert}_T(q_0^2) \right) 
\left( \frac{\alpha^{\rm pert}_T(\mu^2)}{\alpha^{\rm pert}_T(q_0^2)}
\right)^{1-\gamma_0^{A^2}/\beta_0} 
\frac{g^2_T(q_0^2) \langle A^2 \rangle_{R,q_0^2}} {4 (N_C^2-1)}
\right) \ , \nonumber \\
\eeq
where $R(\alpha,\alpha_0)$, $\gamma_0^{A^2}$ and $\alpha_T^{\rm pert}$ can be obtained 
in perturbation and through the OPE analysis (see ref.~\cite{Blossier:2011tf}). 
This can be in Fig.~\ref{fig:plotalphaTPhi}, borrowed from ~\cite{Blossier:2011tf}.
The left plot shows the lattice data for the Taylor coupling multiplied by the square of the momentum plotted 
in terms of the four-loop perturbative value of the coupling at the same momentum, with a best-fit of 
$\Lambda_{\msbar}$ which gives 
\beq
\alpha_{\msbar}(m_Z) \ = \ 0.1198(9) \ , 
\eeq
in very good agreement with PDG~\cite{Nakamura:2010zzi}. One should notice that the departure from zero for the lattice data 
in the plot can be only explained by non-perturbative contributions and that 
the Wilson coefficient for the Landau-gauge gluon condensate 
successfully describes the non-flat behaviour from the data up to 
$\alpha^{\rm pert}\simeq 0.32$ (the solid line given by \eq{alphahNP}), 
which roughly corresponds to $p \simeq 3.5$ GeV. 
The right plot shows the best-fit of \eq{alphahNP} 
to the lattice data, with a best-fit value of 
\beq
g^2(q_0^2) \langle A^2 \rangle_{R,q_0^2} \ = \ 4.5(4) \ \mbox{\rm GeV}^2 \ , 
\eeq
where $q_0^2=100$ GeV$^2$. These are of course estimates for $\Lambda_{\msbar}$ and the gluon condensate 
including 2 degenerate light quarks and two non-degenerates heavier ones. 
This allows to run in perturbation the coupling value up to cross safely the bottom mass threshold 
and run again up to get a ``physical'' estimate for the coupling at the $Z^0$-mass scale, that can 
be directly (and successfully) compared with experimental results. 
The same analysis had been previously applied to lattice data in pure Yang-Mills QCD ($N_f=0$)~\cite{Boucaud:2008gn} 
and with $N_f=2$ twisted-mass dynamical flavours \cite{Blossier:2010ky}, providing with estimates for 
the gluon condensate slightly depending on the number of light quark flavours, although roughly 
ranging from 4-6 GeV$^2$, when statistical uncertainties are included.

\vskip 0.5 cm

\section{The OPE for the ghost-gluon vertex}
\label{sec:OPEsym}

We will now pay attention to (and describe with more details) 
the OPE procedure applied in ref.~\cite{Boucaud:2011eh} for the Landau-gauge ghost-gluon vertex,
\beq
V^{abc}_\mu(-q,k;q-k) &=&
\Gamma^{a'b'c'}_{\mu'}(-q,k;q-k) \ G^{bb'}_{\mu\mu'}(q-k) \ F^{aa'}(q) \
F^{cc'}(k)
\nonumber \\
&=&
\int d^4y \ d^4x \ e^{i (q-k) \cdot x} \ e^{i k \cdot y}
\Braket{ \ T\left( c^c(y) A^b_\mu(x) \overline{c}^a(0) \right)}
\eeq
is quite similar to the one described in ref.~\cite{Boucaud:2008gn}.
Here, the OPE expansion shall read

\beq\label{OPEvertexS}
V^{abc}_\mu(-q,k;q-k) &=&
\left(d_0\right)^{abc}_{\mu}(q,k) \
\nonumber \\
&+&
\left(d_2\right)^{abc\mu'\nu'}_{\mu a'b'}(q,k) \
\Braket{ \  :  A_{\mu'}^{a'}(0) A_{\nu'}^{b'}(0) :}
\ + \ \cdots
\eeq
where $d_0$ accounts for the purely perturbative contribution
to the vertex, while
\beq
w^{a b c}_{\mu} &=&
\left(d_2\right)^{abc\mu'\nu'}_{\mu a'b'}(q,k) \ \delta^{a'b'} g_{\mu'\nu'}
\nonumber \\
&=& 2 I^{[1]} + 2 I_s^{[1]} + 2 I^{[2]} + 4 I^{[3]} +2I^{[4]} + I^{[5]}
\eeq
where\footnote{$N_C$ is the number of colours.}
\beq\label{I1S}
I^{[1]} & = & \ghThreeOneRS \nonumber
\\
%
&=&   \rule[0cm]{0cm}{0.9cm}
\frac{N_C}{2} \ g^2 \
\frac{(q-k)_\sigma q_\sigma}{q^2 (q-k)^2} \
V^{abc}_{{\rm tree}, \mu}(-q,k;q-k) \ ,
\eeq
the tree-level ghost-gluon vertex being
\beq
V^{abc}_{{\rm tree}, \mu}(-q,k;q-k) \ = \ - i \frac{g}{k^2 q^2 (q-k)^2} \ f^{abc} \
q_{\mu_2} \ g^{\perp}_{\mu_2 \mu}(q-k) \ ,
\eeq
while
\beq\label{I1vertexSs}
I_s^{[1]} &=&
\ghThreeOneRSs \ = \
I^{[1]}
\left\{
\begin{array}{c}
q \to -k
\\ k \to -q
\end{array}
\right\} \nonumber  \\
&=&
\altura{0.9}
- \frac{N_C}{2} \ g^2 \
\frac{(q-k)_\sigma k_\sigma}{k^2 (q-k)^2} \
V^{abc}_{{\rm tree}, \mu}(-q,k;q-k) \,,
\eeq
and
\beq\label{I2S}
I^{[2]} \altura{1.8} & = &  \ghThreeTwo \nonumber  \\
&=&   \altura{0.8}
\frac {N_C} {4} g^2 \frac {k_\sigma q_\sigma}{q^2 k^2} \ V_{{\rm tree},\mu}^{a b c}(-q,k;q-k) \  \ .
\eeq
The other contributions, $I^{[3,4,5]}$, can be immediately written using the previous results:

\beq\label{I3S}
I^{[3]} \ &=& \frac 1 2 \left( \ \ghThreeThree + \ghThreeThreeRS \ \right) \nonumber \\
&=& \left( \frac {N_C}{4} \frac{g^2}{k^2} + \ \frac {N_C} {4} \frac{g^2}{q^2} \right) V_{{\rm tree},\mu}^{a b c}(-q,k;q-k)
\eeq
and
\beq\label{I45S} \altura{1.5}
2I^{[4]} +  I^{[5]} \ &=& \ 2\times \ghThreeFour \ +  \ghThreeFive
\nonumber \\
&=&  \ N_C \frac{g^2}{(q-k)^2} \ V_{{\rm tree},\mu}^{a b c}(-q,k;q-k)
\eeq
In all cases, the blue bubble refers to a contraction of the colour and Lorentz indices with $\frac{1}{2}\delta_{ab}g_{\mu\nu}$. We have also introduced the notation $g^\perp_{\mu\nu}(\ell)=g_{\mu\nu}-\frac{\ell_\mu \ell_\nu}{\ell^2}$ for the transverse projector.\\

Then, one obtains
\beq
w^{abc}_\mu \ = \ g^2 \ \left( \altura{0.8} s_V(q,k) + s_F(k) + s_F(q) + s_G(k-q) \right)\,
V_{{\rm tree},\mu}^{a b c}(-q,k;q-k)
\eeq
with
\beq\label{eq:sVsG}
s_G(q) &=& s_F(q) \ = \frac{N_C}{q^2}\,, \nonumber \\
s_V(q,k) &=& \frac {N_C} {2} \left( 2 \ \frac{(q-k)\cdot q}{q^2 (q-k)^2} + 2 \ \frac{(k-q)\cdot k}{k^2 (q-k)^2} + \frac{k\cdot q}{k^2 q^2} \right)  \,.
\eeq
$s_F$ (resp. $s_G$)  comes from the OPE corrections to the external ghost (resp. gluon)
propagator, {\sl i.e.}  from the non-proper diagrams in Eqs.~(\ref{I3S},\ref{I45S})
and $s_V$ from the proper vertex correction in Eqs.~(\ref{I1S},\ref{I2S}).
The OPE corrections to the proper vertex can be obtained from there, remembering that
the bare ghost-gluon vertex reads as:
\beq\label{vertMinko2}
\Gamma^{abc}_{\mu}(-q,k;q-k) \ = \ - g_0 f^{abc}  \
\left( \altura{0.6} q_\mu H_1(q,k) + (q-k)_\mu H_2(q,k) \right) \ .
\eeq
Then, in Landau gauge, only the form factor $H_1$ survives in the Green function to give:
\beq
V^{abc}_{\mu}(-q,k;q-k) &=&
-i g_0 f^{abc} q_{\mu'} g^{\perp}_{\mu'\mu}(q-k) H_1(q,k) \ G((q-k)^2) \ F(q^2) \ F(k^2) \ ,
\eeq
where $G$ and $F$ are the gluon and ghost dressing functions for which we previously computed
the non-perturbative OPE corrections. Thus, one would have:
\beq\label{eq:H1OPE}
H_1(q,k) \ = \ H_1^{\rm pert}(q,k) \left( 1 + s_V(q,k) \ \frac{\braket{A^2}}{4 (N_C^2-1)}
+  \  {\cal O}(g^4,q^{-4}, k^{-4},q^{-2}k^{-2})\ \right).
\eeq
All the non-proper corrections have been removed in the usual way. 
Now, in order to compare all over the available range 
of momenta with some current lattice data, we need to try to 
model the low-momentum behaviour, non-accessible via our previous OPE 
analysis. This is also done in ref.~\cite{Boucaud:2011eh} and will be 
described in the next section.

\section{The Euclidean ghost-gluon vertex and the present lattice data}

We will now devote this section to model the transverse form factor of the Euclidean ghost-gluon
vertex, $H_1$, on the ground provided by \eq{eq:H1OPE}, and compare the result with some recent
lattice data for this form factor computed in different kinematical configurations.

\subsection{The model for the Euclidean ghost-gluon vertex}

In Euclidean metrics the bare ghost-gluon vertex can be written  very similarly  as:
\beq
\Gamma^{abc}_{{\rm bare},\mu}(-q,k;q-k) \ = \ i g_0 f^{abc}  \
\left( \altura{0.6} q_\mu H_1(q,k) + (q-k)_\mu H_2(q,k) \right) \,,
\eeq
where the form factor $H_1$ plays a crucial r\^ole when solving
the ghost-propagator Dyson-Schwinger equation (GPDSE) as
discussed above.
The OPE non-perturbative corrections to the form
factor $H_1$ were obtained in \eq{eq:H1OPE}, but that result
is in principle only reliable for large enough $k,q$ and $q-k$ since the  SVZ factorization
on which it relies might not be valid for low  momenta. Nevertheless, in the following we will propose
a very simple conjecture to extend \eq{eq:H1OPE} to any momenta, which will provide us with
a calculational model for the ghost-gluon vertex to continue research with.
In particular, the perturbative part of the ghost-gluon vertex is usually approximated by
a constant behaviour\footnote{At least for large momenta, this seems to be the case in
lattice simulations~\cite{Cucchieri:2008qm,Sternbeck:2005re} for several kinematical configurations, and
so is confirmed also by the perturbative calculations in Refs~\cite{Chetyrkin:2000dq,Davydychev:1996pb}.}; if we then apply a
finite renormalization prescription such that
\beq
\widetilde{Z}_1(\mu^2) \left. H_1(q,k) \right|_{\mu^2} \ = \ 1 \ ,
\eeq
where the renormalization momentum, $\mu^2$, for a given kinematical configuration
(for instance, $q-k=0$ and $q^2=k^2=\mu^2$) is chosen to be large enough,
on the basis of \eq{eq:H1OPE}, we can conjecture that
\beq\label{eq:H1model}
H_1(q^2,k^2,\theta) &=&
\widetilde{Z}_1^{-1} \ \left[ \altura{0.95} 1 \ + \
\frac{ N_C g^2 \braket{A^2}}{8 (N_C^2-1)} \right. \nonumber \\
&\times&
\left( \frac{\sqrt{k^2q^2} \cos\theta}{k^2 q^2 + m_{\rm IR}^4}
+ \ 2 \ \frac{q^2-\sqrt{k^2q^2} \cos\theta}{q^2 (q^2+k^2 - 2\sqrt{q^2k^2}\cos\theta) + m_{\rm IR}^4}
\right.
\nonumber \\
&&
\left. \left.
+ \ 2 \ \frac{k^2-\sqrt{k^2q^2} \cos\theta}{k^2 (q^2+k^2 - 2\sqrt{q^2k^2}\cos\theta) + m_{\rm IR}^4}
 \right)
\altura{0.95}
\right] \ ,
\eeq
gives a reasonable description  of the ghost-gluon form factor $H_1$ all over the range
of its momenta $q$ and $k$, where $\theta$ stands for the angle between them. The purpose of
\eq{eq:H1model} is to keep the main features of the momentum behaviour of the ghost-gluon
form factor provided by the OPE analysis and to give a rough description of the deep IR
only with the introduction of some IR mass scale, $m_{\rm IR}$, that can be thought as
some sort of IR regulator (related to some kind of effective gluon mass) 
and which is mainly aimed to avoid the spurious singularities resulting from the 
OPE expansion in momentum inverse powers.

Of course, $\mu^2$ being large enough, $\widetilde{Z}_1$ can be
approximated by some constant value for any $\mu^2$ because the logarithmic behaviour
of the ghost-gluon in perturbation theory has been proven to be very
smooth~\cite{Chetyrkin:2000dq,Davydychev:1996pb}. Thus
\eq{eq:H1model} provides us with a very economical model because one
needs nothing but some infrared mass parameter which, being related to the gluon mass,
is expected to be $\sim 1$ GeV, to parametrize the deep infrared behaviour of
the ghost-gluon transverse form factor.
It should be noted that, although the ghost-gluon vertex form factors do not diverge, the
gluon condensate does: it is the product of the Wilson coefficient and of the condensate which
is expected to remain finite thanks to a delicate compensation of singularities,
in the sum-rules approach. Then, both the Wilson coefficient and the condensate should be renormalized,
by fixing some particular prescription, and both shall depend on a chosen renormalization momentum.
This is  explained in detail in the work of ref.~\cite{Blossier:2010vt} where the Wilson coefficient for the
quark propagator is included at  order ${\cal O}(\alpha^4)$.
Since we carry out the computation of the Wilson coefficient in Eqs.~(\ref{I1S}-\ref{I45S}) only at  tree-level
and neglect its logarithmic dependence, we do not specify any renormalization momentum
for the gluon condensate.

\subsection{Comparing with available lattice data}

\begin{figure}[h]
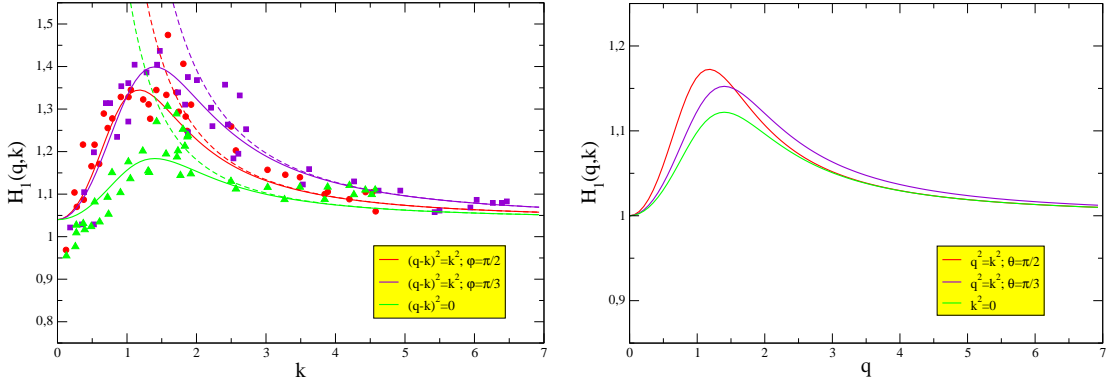

\begin{center}
\begin{tabular}{cc}
\includegraphics[width=7.2cm]{figs/H1SU2.eps} &
\includegraphics[width=7cm,height=5cm]{figs/H1SU3.eps}
\end{tabular}
\end{center}
\caption{\small (left) The predictions of the model for $H_1$ (solid lines) in Eq.~(4.3) 
and the OPE result without infrared completion (dotted lines) in Eq.~(3.14) 
confronted to the SU(2) lattice data borrowed from ref.~\cite{Cucchieri:2008qm} for the three kinematical
configurations described in the main text: $(k-q)^2=q^2$ for $\varphi=\pi/3$ (violet)
and $\varphi=\pi/2$ (red) and $(k-q)^2=0$ (green).
(right) The results of the model for $H_1$ in the SU(3) case and for the following kinematical configurations:
$k^2=q^2$ for $\theta=\pi/3$ (violet) and $\theta=\pi/2$ (red) and $k^2=0$ (green).}
\label{fig:lat}
\end{figure}

Some  SU(2) and SU(3) lattice results for the ghost-gluon vertex are available in the literature (see for instance
Refs.~\cite{Cucchieri:2008qm,Sternbeck:2005re}). In particular, Cucchieri et al.   have computed the tree-level-tensor
form factor in SU(2) for three kinematics  ($p \equiv q - k$ is the gluon momentum and  $\varphi$ the angle
between the gluon and ghost momenta) : (1) $p^2=q^2$ and $\varphi=\pi/2$,  (2)  $p^2=0$
and (3) $\varphi=\pi/3$ with $p^2=q^2=k^2$ (\cite{Cucchieri:2008qm}).
As can be seen in the left pane of Fig.~\ref{fig:lat},
 when we plug $g^2 \langle A^2\rangle=6$ GeV$^2$, $m_{\rm IR}=1.4$ GeV  and $\widetilde{Z}_1^{-1}=1.04$
 into \eq{eq:H1model}, the prediction for the form factor $H_1$ appears to agree pretty well
 with the SU(2) lattice data for the three kinematical configurations above mentioned. It should be also noted 
 that the OPE prediction without the infrared completion through the introduction of $m_{\rm IR}$ 
 given by Eqs.~(\ref{eq:sVsG},\ref{eq:H1OPE}), plotted with dotted lines, also accounts very well for 
 lattice data in the intermediate momenta region, above roughly $2.5$ GeV.  
 Furthermore, the
 value for the gluon condensate lies in the same ballpark as the estimates\footnote{In ref.~\cite{Boucaud:2008gn},
 $g^2 \langle A^2\rangle$ is evaluated through an OPE formula with a tree-level Wilson coefficient to be 5.1 GeV$^2$
 at a renormalization point of 10 GeV.} obtained from the SU(3) analysis of the Taylor coupling
 in ref.~\cite{Boucaud:2008gn}. This strongly supports that the OPE analysis indeed captures the kinematical 
 structure for the form factor $H_1$.  
 On the other hand, the infrared mass scale, as supposed, is of the order of 1 GeV
 and the (very close to 1) value for $\widetilde{Z}_1^{-1}$ accounts reasonably for perturbative
 value of $H_1$ that we approximated by a constant. Thus, after paying the economical price of 
 incorporating only these two more parameters, both taking also reasonable values, we are left with 
 a reliable closed formula for the form factor at any gluon or ghost momenta. This is a very 
 useful ingredient for the numerical integration of the GPDSE.

The SU(3) results published by the authors of ref.~\cite{Sternbeck:2005re} appear to be
very noisy and cannot be invoked to properly discriminate whether a constant behaviour close to 1
or \eq{eq:H1model} accounts better for them.
However,  we can  now take the mass parameter, $m_{\rm IR}$, from the
previous SU(2) analysis and the well-known SU(3) value for $g^2 \braket{A^2}$,
assume $\widetilde{Z}_1^{-1}=1$ and use \eq{eq:H1model} to predict
the ghost-gluon transverse form factor, $H_1$. This is shown in the right plot of in Fig.~\ref{fig:lat} for
three different kinematical configurations and, as can be seen, the deviations from 1
appear to be very small in all the cases and compatible with the results shown in
Fig.~4 of ref.~\cite{Sternbeck:2005re} for the vanishing gluon momentum case.


\subsection{The Taylor kinematics and the asymmetric gluon-ghost vertex}

The kinematical configuarition where the incoming ghost
momentum goes to zero ($k \to 0$) is especially interesting, as this is the one 
which one defines the T-scheme for (as shown in sec.\ref{sec:VA}).  
Let us pay some additional attention to it. In this particular kinematical limit, 
it is exactly obtained up to all
perturbative\footnote{The same argument of Taylor's perturbative proof still works
if one consider the Landau-gauge ghost-gluon vertex DSE: a vanishing ghost momentum entering in the vertex
implies the contraction of the gluon-momentum transversal projector with the gluon momentum itself for any dressed
diagram. Thus, Taylor's theorem is still in order within the non-perturbative DSE framework. Some
attention have been recently paid to the ghost-gluon vertex DSE~\cite{Schleifenbaum:2004id}.} orders:
\beq \label{taylor}
\Gamma^{abc}_{{\rm bare},\mu}(-q,0;q) \ = \ - g f^{abc} q_\mu \ .
\eeq
At any order of perturbation theory, this implies that $H_1(q,0)+H_2(q,0)=1$. According to \eq{eq:H1model}, one would have
\beq\label{H1OPETaylor}
H_1(q,0) \ = \  \widetilde{Z}_1^{-1} \
\left( 1 +  N_C \frac{g^2 \braket{A^2}}{4 (N_C^2-1)} \ \frac {q^2} {q^4+m_{IR}^4}  \ \right)
\ .
\eeq
An interesting question to investigate is
whether such a genuine non-perturbative correction still survives for the full ghost-gluon
vertex in the Taylor limit, and not only for the transverse form factor $H_1$. Said otherwise,
does eq.\eqref{taylor} maintains its validity upon adding the OPE corrections related to $\VA$ to it, or not?
This question have been properly addressed in ref.~\cite{Boucaud:2011eh}, where 
the asymmetric ghost-gluon vertex is studied and one is left with
\beq\label{OPEamp2}
\Gamma^{abc}_\mu(-q,\varepsilon;q-\varepsilon)  \ = \
- g f^{abc} \left( \altura{0.6} q_\mu \ H_1(q,\varepsilon) \ + \
(q-\varepsilon)_\mu \ H_2(q,\varepsilon) \right)
\ + \ \cdots
\eeq
where
\beq
H_1(q,\varepsilon) &=&  H_1^{\rm pert}(q,\varepsilon) \ + \
N_C \ g^2 \frac{(q-\varepsilon) \cdot q}{q^2 (q-\varepsilon)^2} \
\frac{\braket{A^2}}{4 (N^2_C-1)} \ , \nonumber \\
H_2(q,\varepsilon) &=& H_2^{\rm pert}(q,\varepsilon) \ - \
N_C \ g^2 \ \frac{\left((q-\varepsilon)\cdot q \right)^2}{q^2 (q-\varepsilon)^4} \
\frac{\braket{A^2}}{4 (N^2_C-1)} \ .
\label{H12OPE}
\eeq
Thus, after taking the limit $\varepsilon \to 0$, one will have:
\beq
\Gamma^{abc}_\mu(-q,0;q)  \ = \
- g f^{abc} \ q_\mu \left( \ \altura{0.6} H_1^{\rm pert}(q,0) \ + \ H_2^{\rm pert}(q,0) \ \right)
\ = \ - g f^{abc} \ q_\mu \ .
\eeq
Thus, the result applied in \eq{alphahNP} to define the T-scheme coupling is recovered. 
This confirms that no non-perturbative OPE correction survives in
the proper ghost-gluon vertex in the Taylor limit, although both form factors $H_1$ and $H_2$ separately 
undergo such a kind of correction.


\section{Conclusions}

The momentum behaviour of the ghost-gluon vertex has been studied in the framework of the operator product expansion (OPE).
This approach has already been proved in the literature to be very fruitful in describing the running of two-point gluon and quark Green functions
and of the strong coupling computed in several MOM renormalization schemes. In particular, the coupling derived from
the ghost-gluon vertex renormalized in T-scheme, which can be directly computed from nothing else but the ghost and gluon
propagators, led to a very accurate determination of $\Lambda_{\rm QCD}$ when its running with momenta obtained from
the lattice estimate has been confronted with the OPE prediction, although only after accounting properly for a dimension-two gluon
condensate, $\braket{A^2}$. The same approach is followed to study the ghost-gluon vertex
and special attention is payed to the transverse form factor for the ghost-gluon vertex, precisely the one that plays a crucial
r\^ole for the truncation and resolution of the GPDSE and that, in this context, is usually approximated by a constant.
In this way, a genuine non-perturbative OPE correction to the perturbative part for the transverse form
factor is obtained. A very simple conjecture, made to extend the OPE description beyond the momentum range where the SVZ factorization
is supposed to work, left us with a simple model describing the momentum behaviour of the ghost-gluon form factor.
This model was shown to agree pretty well with a lattice SU(2) computation for several kinematical
configurations of the ghost-gluon vertex, when the value for the gluon condensate is in the same ballpark as the one that is estimated
from the running of the T-scheme coupling in SU(3) lattice gauge theory. This successful comparison with the available
lattice data provides us with strong indications that (i) the OPE framework helps to account for the
kinematical structure of the ghost-gluon vertex at intermediate momentum domain
and (ii) that it helps to cook up a reliable model, providing a good parameterization also for the IR domain,
to be plugged into the GPDSE to reproduce the ghost propagator lattice data.
Finally, we also proved that, in the particular kinematical limit associated with the Taylor theorem and hence by the
T-scheme (a vanishing incoming ghost momentum), the corrections to the longitudinal and transverse form factors cancel against
each other and  that the tree-level result for the ghost-gluon vertex is thus recovered.

\bigskip

{\bf Acknowledgements:}
 J. R-Q acknowledges the Spanish MICINN for the support by the research project FPA2009-10773 and ``Junta de Andalucia''
by P07FQM02962. 



\providecommand{\href}[2]{#2}\begingroup\raggedright
\endgroup

\end{document}